\documentclass[showpacs,preprintnumbers,twocolumn,prd]{revtex4}
\usepackage{amsmath,amssymb}
\usepackage[dvips]{graphicx}

\newcommand{\MeV}{\;\text{MeV}}
\newcommand{\GeV}{\;\text{GeV}}

\newcommand{\tr}{\mathrm{tr}}
\newcommand{\rmd}{\mathrm{d}}
\newcommand{\rme}{\mathrm{e}}
\newcommand{\rmi}{\mathrm{i}}
\newcommand{\gs}{g_{\text{S}}}
\newcommand{\gd}{g_{\text{D}}}
\newcommand{\ellb}{\bar{\ell}}

\begin{document}
\title{Isentropic thermodynamics in the PNJL model}
\author{Kenji Fukushima}
\affiliation{Yukawa Institute for Theoretical Physics,
 Kyoto University, Kyoto 606-8502, Japan}
\begin{abstract}
 We discuss the isentropic trajectories on the QCD phase diagram in
 the temperature and the quark chemical potential plane using the
 Nambu--Jona-Lasinio model with the Polyakov loop coupling (PNJL
 model).  We impose a constraint on the strange quark chemical
 potential so that the strange quark density is zero, which is the
 case in the ultra relativistic heavy-ion collisions.  We compare our
 numerical results with the truncated estimates by the Taylor
 expansion in terms of the chemical potential to quantify the
 reliability of the expansion used in the lattice QCD simulation.  We
 finally discuss the strange quark chemical potential induced by the
 strangeness neutrality condition and relate it to the ratio of the
 Polyakov loop and the anti-Polyakov loop.
\end{abstract}
\preprint{YITP-08-84}
\maketitle


\section{INTRODUCTION}

Thermodynamic properties of hot and dense matter out of quarks and
gluons are of theoretical and experimental importance.  To this end
the Monte-Carlo method on the lattice has worked quite successfully to
simulate the matter at high temperature from the first principle, that
is, Quantum Chromodynamics
(QCD)~\cite{Bernard:1996cs,Aoki:2005vt,Aoki:2006br,Cheng:2007jq,DeTar:2008qi}.

It is believed that such hot and dense matter has been created in the
Relativistic Heavy Ion Collider (RHIC) located at BNL.\ \ Since the
baryon stopping power in the nucleus-nucleus collision is small at
the RHIC (top) energy $\sqrt{s_{_{NN}}}=200\GeV$, the created matter is
nearly free from the net baryon density, and thus, the corresponding
baryon chemical potential is much smaller than the temperature.

We are facing two experimental possibilities for the future
exploration of the QCD phase diagram:  One is going toward even hotter
matter as planned in the Large Hadron Collider (LHC) at CERN.\ \ The
other one is realizing denser matter, which is accessible by
collisions at smaller energy with larger baryon stopping power.  The
formation of baryon-rich matter is within the scope of the Facility
for Antiproton and Ion Research (FAIR) at GSI and the systematic
energy scan is also under discussion in the future plan for RHIC with
emphasis on the QCD critical point search.  The present work is
focused on the latter; the thermodynamic properties of matter with a
finite quark chemical potential $\mu$ whose magnitude is comparable to
the temperature $T$.

The lattice QCD simulation has a serious limitation if a finite
chemical potential is turned on in the Dirac
operator~\cite{Schmidt:2008cf,Ejiri:2008nv}.  That is, the notorious
sign problem prevents us from applying the Monte-Carlo method to a
finite-density system.  There are a number of proposals to tame this
problem~\cite{Muroya:2003qs}, among which the Taylor expansion in
terms of $\mu/T$ seems to work well insofar as $\mu/T$ is under the
radius of
convergence~\cite{Allton:2005gk,Bernard:2007nm,Gavai:2008zr}.  In this
way the equation of state (EoS) at finite $T$ and $\mu$ is partially
available from the direct lattice QCD simulation.  Actually the EoS is
an indispensable input for the sake of the hydrodynamic evolution of
matter.  From the $s/n_B$-constant line, where $s$ is the entropy
density and $n_B$ is the baryon number density, we can draw the
isentropic trajectory along which the adiabatic system
evolves~\cite{Bernard:2007nm,Ejiri:2005uv,Miao:2008sz}.

An interesting progress has been made in Ref.~\cite{Bernard:2007nm};
the isentropic thermodynamics was investigated with a constraint that
the strange quark density must be zero, i.e.\ $n_s=0$.  Such a
constraint is necessary to emulate matter created by the ultra
relativistic heavy-ion collisions because thermalization is achieved
within the time scale of the strong interaction and the system is far
from $\beta$-equilibration.

In this paper we will utilize the Nambu--Jona-Lasinio model with the
Polyakov loop coupling (PNJL
model)~\cite{Gocksch:1984yk,Ilgenfritz:1984ff,Fukushima:2003fw,%
Ratti:2005jh,Kashiwa:2007hw,Fu:2007xc,Ciminale:2007sr,Fukushima:2008wg}
to reveal the isentropic
thermodynamics with $n_s=0$ imposed.  In contrast to the NJL model
study along the same line~\cite{Scavenius:2000qd} (see also
Ref.~\cite{Bluhm:2007nu} for another type of approach), the PNJL model
has an advantage that a part of the gluon degrees of freedom is
included, which gives us a hope that the PNJL model can lead to a
better EoS reflecting the gluodynamics as well as the chiral dynamics.
In Ref.~\cite{Kahara:2008yg} we can already find the isentropic
trajectory evaluated in the PNJL model and notice that they are
qualitatively similar to the results in Ref.~\cite{Scavenius:2000qd}.
We will see that the constraint $n_s=0$ causes only little change in
the resultant isentropic trajectory because the strange quark mass
$M_s$ is still substantially large around the crossover region and the
strange density is suppressed in any case.

The extension from the NJL model to the PNJL model is not a minor
improvement, however.  Our nontrivial finding is, in fact, that the
PNJL model is capable of capturing the correct behavior of the induced
strange quark chemical potential $\mu_s>0$ to keep $n_s=0$.  Since the
nonzero $\mu_s$ has its origin in confinement physics, as we will
discuss later, the NJL model without any confinement effect is of no
use but the PNJL model naturally provides us with $\mu_s>0$ which is
related to the ratio of the Polyakov loop $\ell$ and the anti-Polyakov
loop $\bar{\ell}$.  As a matter of fact, although it is known that
$\ell$ deviates from $\bar{\ell}$ at
$\mu\neq0$~\cite{Allton:2005gk,Ratti:2005jh,Dumitru:2005ng,Fukushima:2006uv},
our present work is the very first demonstration to show that this
difference $\ell\neq\bar{\ell}$ has a physical consequence in the best
of our knowledge.

This paper is organized as follows:  We explain the model definitions
in Sec.~\ref{sec:model}.  Then, in Sec.~\ref{sec:results}, we show the
numerical results with the constraint $n_s=0$, followed by the
discussions on the validity of the Taylor expansion in
Sec.~\ref{sec:Taylor}.  We elucidate the physical meaning of
$\mu_s\neq0$ in Sec.~\ref{sec:mus} and the summary is in
Sec.~\ref{sec:summary}.


\section{MODEL SETUP}
\label{sec:model}

The thermodynamic potential of the three-flavor PNJL model consists of
three pieces,
$\Omega=\Omega_{\text{vacuum}}+\Omega_{\text{thermal}}+\Omega_{\text{Polyakov}}$,
namely, the vacuum part (zero-point energy and the condensation
energy), the thermal part, and the Polyakov loop potential,
respectively.  We shall take a close look at them in order.

The vacuum part is exactly the same as in the ordinary NJL model;
\begin{equation}
 \begin{split}
 & \Omega_{\text{vacuum}} = -6\sum_i \int^\Lambda
  \!\frac{\rmd^3p}{(2\pi)^3} \;\varepsilon_i(p) \\
 & + \gs \Bigl( \langle\bar{u}u\rangle^2
  \!+ \langle\bar{d}d\rangle^2 \!+\langle\bar{s}s\rangle^2 \Bigr)
  + 4\gd \langle\bar{u}u\rangle
  \langle\bar{d}d\rangle\langle\bar{s}s\rangle \,,
 \end{split}
\end{equation}
where $i$ refers to the quark flavor running over $u$, $d$, and $s$.
The energy dispersion relations are
$\varepsilon_i(p)=\sqrt{p^2+M_i^2}$ with
$M_i=m_i-2\gs\langle\bar{q}_i q_i\rangle-2\gd\epsilon_{ijk}
\langle\bar{q}_j q_j\rangle\langle\bar{q}_k q_k\rangle$, where $\gs$
represents the four-Fermi coupling constant and $\gd$ represents the
't~Hooft interaction strength.  There are three more model parameters:
the light current quark mass $m_u=m_d$, the heavy current quark mass
$m_s$, and the ultraviolet cutoff $\Lambda$.  We here adopt the
parameter set by Hatsuda-Kunihiro~\cite{Hatsuda:1994pi};
\begin{equation}
 \begin{split}
 &\Lambda = 631.4\MeV \,, \\
 &\gs\Lambda^2 = 3.67 \,, \qquad\qquad\quad\!
  \gd\Lambda^2 = -9.29 \,, \\
 &m_u = m_d = 5.5\MeV \,, \quad
  m_s = 135.7\MeV \,,
 \end{split}
\end{equation}
which fits $m_\pi$, $m_K$, $m_{\eta'}$, $f_\pi$, and empirical $M_{ud}$.

The thermal part has a coupling to the Polyakov loop (spatially
homogeneous $A_4$ background) in a form as
\begin{align}
 \Omega_{\text{thermal}} &= -2T\sum_i \int\frac{\rmd^3p}{(2\pi)^3}
  \;\Bigl\{\ln\det\Bigl[1+L\rme^{-(\varepsilon_i(p)-\mu)/T}\Bigr]
  \notag\\
  & \qquad +\ln\det\Bigl[1+L^\dagger\rme^{-(\varepsilon_i(p)+\mu)/T}
  \Bigr]\Bigr\} \,,
\end{align}
where the Polyakov loop is a $3\times3$ matrix in the fundamental
representation in color space defined by
\begin{equation}
 L(\vec{x}) = \mathcal{P}\exp\biggl[-\rmi g\int_0^\beta \rmd x_4\,
  A_4(\vec{x},x_4)\biggr] \,,
\end{equation}
and in this paper the Polyakov loop sometimes refers to the traced
one after average, that is,
\begin{equation}
 \ell = \frac{1}{3}\bigl\langle\tr L\bigr\rangle \,,\qquad
 \bar{\ell} = \frac{1}{3}\bigl\langle\tr L^\dagger\bigr\rangle \,,
\end{equation}
as long as no confusion may arise.  In a simple mean-field
approximation for the Polyakov loop (group) integration, the
determinant explicitly reads;
\begin{align}
 & \det\Bigl[1+L\rme^{-(\varepsilon-\mu)/T}\Bigr] \to 1
  +\rme^{-3(\varepsilon-\mu)/T} \notag\\
 &\qquad\qquad\qquad +3\,\ell\,\rme^{-(\varepsilon-\mu)/T}
  +3\,\ellb\,\rme^{-2(\varepsilon-\mu)/T} \,,\\
 & \det\Bigl[1+L^\dagger\rme^{-(\varepsilon+\mu)/T}\Bigr] \to 1
  +\rme^{-3(\varepsilon+\mu)/T} \notag\\
 &\qquad\qquad\qquad + 3\,\ellb\,\rme^{-(\varepsilon+\mu)/T}
  +3\,\ell\,\rme^{-2(\varepsilon+\mu)/T} \,.
\label{eq:determinant}
\end{align}
It is important to note that a positive $\mu$ induces $\ellb>\ell$,
while $\ellb=\ell$ at $\mu=0$.  The finite-temperature field theory
tells us that the traced Polyakov loop gives the exponential of the
free energy cost by a test quark, i.e.\ $\ell=\rme^{-f_q/T}$ up to
normalization (or energy offset), and the anti-Polyakov loop by a test
anti-quark, i.e\ $\ellb=\rme^{-f_{\bar{q}}/T}$~\cite{Svetitsky:1985ye}.
Therefore, $f_q\to\infty$ and thus $\ell\to0$ signifies quark
confinement, so that the Polyakov loop serves as an order parameter
for the color deconfinement phase transition.  In the presence of
dynamical quarks in the color fundamental representation, however,
neither $\ell$ nor $\ellb$ can be strictly zero due to screening.  If
the color screening is stronger, the free energy cost is smaller, and
the Polyakov loop is larger accordingly.  In a medium with $\mu>0$ the
test anti-quark is screened more efficiently than the test quark, that
means
$\ellb>\ell$~\cite{Allton:2005gk,Ratti:2005jh,Dumitru:2005ng,%
Fukushima:2006uv}.

We shall choose the Polyakov loop potential as~\cite{Fukushima:2008wg}
\begin{equation}
 \begin{split}
 & \Omega_{\text{Polyakov}} = -b\cdot T\Bigl\{
  54\,\rme^{-a/T}\,\ell\,\bar{\ell} \\
 & \qquad + \ln\bigl[ 1-6\,\ell\,\bar{\ell}-3(\,\ell\,\bar{\ell}\,)^2
  +4(\,\ell^3 + {\bar{\ell}}^3\,) \bigr] \Bigr\} \,.
 \end{split}
\end{equation}
Here, there are two parameters $a$ and $b$ in the above ansatz.  We
fix $a=664\MeV$ to reproduce $T_c\simeq270\MeV$ in the pure gluonic
sector and $b=0.03$ to yield $T_c\simeq200\MeV$ for the simultaneous
crossovers of deconfinement and chiral restoration.


\section{RESULTS}
\label{sec:results}

We are now ready to proceed to the numerical calculations using the
PNJL model.  Here let us focus on the entropy per baryon density
ratio, $s/n_B$.  This is because the adiabatic hydrodynamic expansion
conserves $s/n_B$ along the time evolution.  It is easy to confirm
that $s/n_B$ is a constant indeed from conservation of the entropy
current and the baryon current, that is,
$(\rmd/\rmd\tau)(s/n_B)=u^\mu\partial_\mu(s/n_B)=0$ readily follows from
$\partial_\mu(s u^\mu)=0$ and $\partial_\mu(n_B u^\mu)=0$.  Hence, one
does not have to integrate the hydrodynamic equation to draw the
time-evolution path, which is simply inferred from an $s/n_B$-constant
line .  This is the case insofar as the expansion is fast enough to
make the system thermally isolated and the entropy production due to
dissipation is negligible.


\subsection{Case without constraint on $n_s$}

We first take a quick look at the results without constraint on $n_s$.
The quark density is specified by the quark chemical potential (or one
third of the baryon chemical potential) which is common to all three
flavors.  In the results presented in this subsection, thus, the net
strange quark density $n_s$ is nonzero.

\begin{figure}
\includegraphics[width=9.5cm]{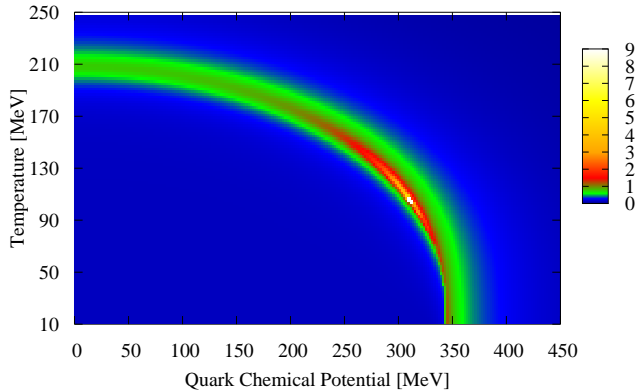}
\caption{Chiral susceptibility with respect to light ($u$ and $d$)
  quarks defined by $-\rmd^2\Omega/\rmd m_u^2$ in the unit of
  $\Lambda$ as a function of $\mu$ and $T$.}
\label{fig:contour}
\end{figure}

We solve the following gap equations self-consistently;
\begin{equation}
 \frac{\partial\Omega}{\partial\langle\bar{u}u\rangle} =
 \frac{\partial\Omega}{\partial\langle\bar{s}s\rangle} =
 \frac{\partial\Omega}{\partial\ell} =
 \frac{\partial\Omega}{\partial\ellb} = 0
\label{eq:gap_eq}
\end{equation}
with assuming isospin symmetry
$\langle\bar{d}d\rangle=\langle\bar{u}u\rangle$.  In this way we have
the chiral condensates, the Polyakov loop, and the anti-Polyakov loop
as functions of $T$ and $\mu$.  To illustrate the phase structure, we
show the light-quark chiral susceptibility in a density plot in
Fig.~\ref{fig:contour}.

We can perceive from Fig.~\ref{fig:contour} that the critical region
extends in the vicinity of the second-order critical point located at
$(T,\,\mu)=(315\MeV,\,100\MeV)$ on top of the enhanced strip along the
chiral crossover.  Because our approximation neglects the
soft-mode fluctuations, the EoS obtained in this work may miss the
singular contribution to thermodynamic quantities near the critical
point~\cite{Nonaka:2004pg,Schaefer:2006ds,kamikado}.  We already know,
however, that the PNJL model can reproduce the pressure behavior at
$\mu=0$, which implies that the soft-mode contribution is not
significant there.  We can then anticipate that the singular
contribution would become important only in a narrow region
surrounding to the critical point (indicated by the bright colors on
the density plot in Fig.~\ref{fig:contour}).  In what follows we shall
draw the isentropic trajectories onto this phase structure not taking
account of soft-mode fluctuations.


\begin{figure}
 \includegraphics[width=8.5cm]{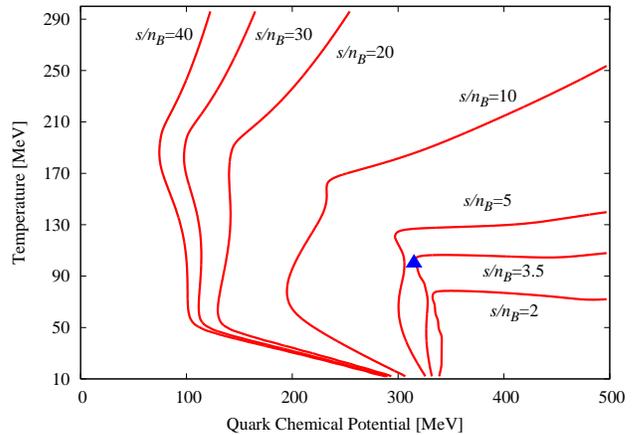}
 \caption{Isentropic trajectories on the $\mu$-$T$ plane in the case
  without constraint on $n_s$.  The (blue) triangle marks the location
  of the critical point.}
\label{fig:isen_traj}
\end{figure}

By substituting the solution of the gap equations into $\Omega$, the
thermodynamic functions such as the entropy and the baryon number of
our current interest obtain by means of the thermodynamic relations;
\begin{equation}
 s = -\frac{\partial\Omega}{\partial T} \,,\qquad
 n_B = -\frac{1}{3}\frac{\partial\Omega}{\partial\mu} \,.
\end{equation}
Figure~\ref{fig:isen_traj} shows the numerical results for the
isentropic trajectories for various values of $s/n_B$ without imposing
constraint on $n_s$.

We note that Fig.~\ref{fig:isen_traj} is reasonably consistent with
Fig.~9 in Ref.~\cite{Kahara:2008yg} which employs the two-flavor PNJL
model.  Moreover, the trajectories look qualitatively similar to the
results in the NJL model as in Ref.~\cite{Scavenius:2000qd}, while the
value of $s/n_B$ associated with each trajectory is greater in our
case than in the NJL model study.  This has an intuitive
interpretation.  The PNJL model is composed of quasi-quarks and a part
of gluons, so that $s$ has steeper behavior near $T_c$ and grows
larger above $T_c$ as compared to the NJL model.  As for $n_B$,
because of the Polyakov loop average, $n_B$ has steeper behavior as
well, but it does not exceed the NJL model value.  The ratio $s/n_B$
in the PNJL model, therefore, results in mild sensitivity to the
steepness of $s$ and $n_B$ near $T_c$, leading to the similar
trajectory curves to the NJL model results.  In contrast, the
magnitude of $s/n_B$ corresponding to the trajectory becomes larger in
the PNJL model as a consequence of the gluon degrees of freedom
contributing to $s$.

Comparing our results to the lattice QCD simulation with the Taylor
expansion~\cite{Miao:2008sz}, we see that our estimate of $s/n_B$ (for
instance comparing the $s/n_B=40$ curve in Fig.~\ref{fig:isen_traj}
and the $s/n_B=45$ curve in Ref.~\cite{Miao:2008sz}) improves an
agreement.


We remark that all the trajectories in the low density side must go to
$\mu>M_u$ as $T$ decreases because $n_B\to0$ when $\mu<M_u$ and
$T\to0$.  In the hadron phase at small $T$ and moderate $\mu$, in any
case, the PNJL model description is not realistic as it cannot
describe the nucleon and nuclear matter.  We should be aware that the
PNJL model works well near and above $T_c$ and $\mu_c$ but not far
below them.



\subsection{Case with constraint on $n_s$}

We next proceed to the case with imposing $n_s=0$ to emulate the
situation in the high-energy heavy-ion collisions.  We should
determine $\mu_s$ self-consistently solving
\begin{equation}
 n_s = -\frac{\partial\Omega}{\partial\mu_s} = 0 \,,
\label{eq:ns_const}
\end{equation}
together with other gap equations in Eq.~(\ref{eq:gap_eq}).

\begin{figure}
\includegraphics[width=8.5cm]{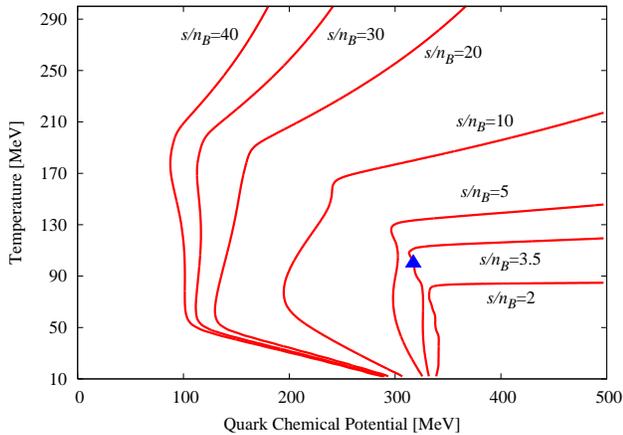}
\caption{Isentropic trajectories with $n_s=0$ imposed, which is
  relevant to the relativistic heavy-ion collisions.  The (blue)
  triangle marks the location of the critical point in this case.}
\label{fig:isen_traj_ns}
\end{figure}

\begin{figure}
\includegraphics[width=8.5cm]{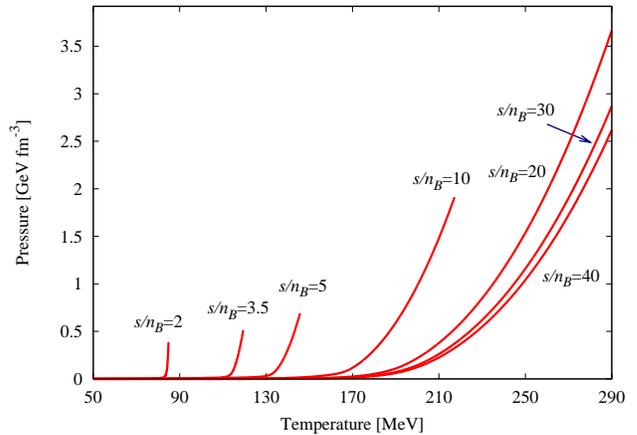}
\caption{Pressure along the respective isentropic trajectories
  given in Fig.~\ref{fig:isen_traj_ns}.}
\label{fig:p_isen_ns}
\end{figure}

The phase structure is only slightly changed by the constraint.  The
critical point moves from $(\mu,T)=(315\MeV,100\MeV)$ to
$(\mu,T)=(317\MeV,100\MeV)$.  The resultant isentropic trajectory as
shown in Fig.~\ref{fig:isen_traj_ns} thus takes a very similar shape
to the case without constraint $n_s=0$.  We further calculate the
pressure along the isentropic trajectories on
Fig.~\ref{fig:isen_traj_ns} and make a plot of
Fig.~\ref{fig:p_isen_ns}.  Here we did not normalize the pressure by
$T^4$ as is often the case, for the density contribution
$\propto\mu^4$ becomes predominant as $s/n_B$ goes small.


From Fig.~\ref{fig:isen_traj_ns} we see that the strangeness
neutrality condition has a noticeable effect on the trajectory only at
high $T$.  This is because the $s$-quark constituent mass, $M_s$, is
still heavy around the chiral crossover with respect to $u$ and $d$
quarks, and so $n_s\approx0$ even without constraint as long as $T$ is
low and $\mu$ is smaller than $M_s$.  If $\mu$ surpasses $M_s$, we
would recognize a difference by the neutrality effect in the low-$T$
region, but then, we have to consider color superconductivity at such
high density, which is not within our current scope.

The isentropic thermodynamics hardly changes, as observed in
Fig.~\ref{fig:p_isen_ns}, until $s/n_B\lesssim20$.  This behavior
agrees with the lattice results~\cite{Bernard:2007nm,Miao:2008sz} in
which no significant $s/n_B$ dependence has been found.

We could have placed a plot here for the induced $\mu_s$ as a function
of $T$.  For later convenience, however, we postpone showing it and
let us turn to the validity of the Taylor expansion as utilized in the
lattice QCD simulation.


\section{VALIDITY OF THE TAYLOR EXPANSION}
\label{sec:Taylor}

All of our calculations result from the PNJL model in the mean-field
approximation and do not rely on the Taylor expansion unlike the
lattice QCD simulation.  Although we have no need to carry the
expansion out, it is interesting to compare ``full'' results of our
numerical calculations and ``truncated'' ones to verify how nicely the
Taylor expansion works.

Within the framework of the PNJL model we have solved $s$ and $n_B$ as
functions of $\mu$.  Now we shall expand them into the series like
Ref.~\cite{Ratti:2007jf} as
\begin{align}
 s(\mu) &= \lim_{M\to\infty}\sum_{n=0}^M
  \frac{\rmd^{2n} s(0)}{\rmd \mu^{2n}}\, \mu^{2n} \,,
\label{eq:ex_s}\\
 n_B(\mu) &= \lim_{N\to\infty}\sum_{n=0}^N \frac{\rmd^{2n+1} n_B(0)}
  {\rmd \mu^{2n+1}}\, \mu^{2n+1} \,,
\label{eq:ex_nb}
\end{align}
which is to be validated if no singularity associated with the
first-order phase transition lies along the $\mu$-direction.  It
should be mentioned that the derivative in Eqs.~(\ref{eq:ex_s}) and
(\ref{eq:ex_nb}) is the total derivative in a sense that it acts on
the implicit $\mu$-dependence in the mean-fields.  In this way the
mixing effect can be correctly taken into account in the model
treatment~\cite{Sasaki:2006ww}.  From symmetry $s$ is an even function
of $\mu$ and $n_B$ is an odd function.

Let us explain the numerical procedure in details to make a comparison
between the results with and without truncation of the degree in the
Taylor expansion.  We first approximate $s(\mu)$ and $n_B(\mu)$ by the
Taylor expansion with sufficiently large number of $M$ and $N$ so that
the coefficients in the first several terms barely change with an
increment of $M$ and $N$.

We should further specify the fit range of $\mu\in[0,\mu_0]$ to read
the Taylor expansion coefficients.  In principle, if the ``exact''
calculation were possible, $\mu_0$ could be zero or there is no $\mu_0$
dependence at all.  Even in the mean-field level, however, we are far
from the exact calculation.  Here we have chosen $\mu_0=50\MeV$ and
$\mu_0=100\MeV$.  In fact, $\mu_0$ is a parameter which controls the
precision in the determination of the Taylor expansion coefficients in
the same sense as using the multiple-point formula for the numerical
differentiation.  One might have thought that $\mu_0=500\MeV$, for
instance, can cover the whole density region in
Fig.~\ref{fig:isen_traj_ns}, but such a choice would bring artifact
from outside the radius of convergence.

Once we fix the Taylor expansion coefficients of $s(\mu)$ and
$n_B(\mu)$, then we cut the series at smaller $M$ and $N$.  We will
elucidate the leading-order case ($M=2$ and $N=1$) and the next to
leading-order case ($M=3$ and $N=2$) below.


\begin{figure}
\includegraphics[width=8.5cm]{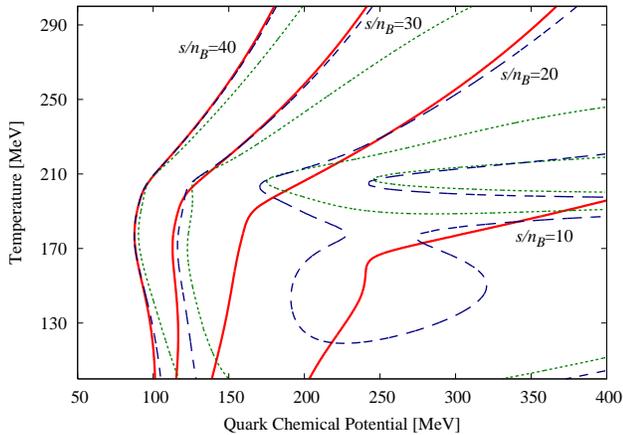}
\caption{Comparison between the full results (solid curve) and the
  truncated Taylor expansion up to the $\mu^4$-term in $s(\mu)$ and
  the $\mu^3$-term in $n_B(\mu)$.  The (green) dotted and (blue)
  dashed curves represent the estimates by the Taylor series fitted to
  reproduce the full results in the ranges from zero to $50\MeV$ and
  $100\MeV$, respectively.  Four curves are for $s/n_B=40$, $30$,
  $20$, $10$ from the left to the right.}
\label{fig:ex_isen_2}
\end{figure}


\begin{figure}
\includegraphics[width=8.5cm]{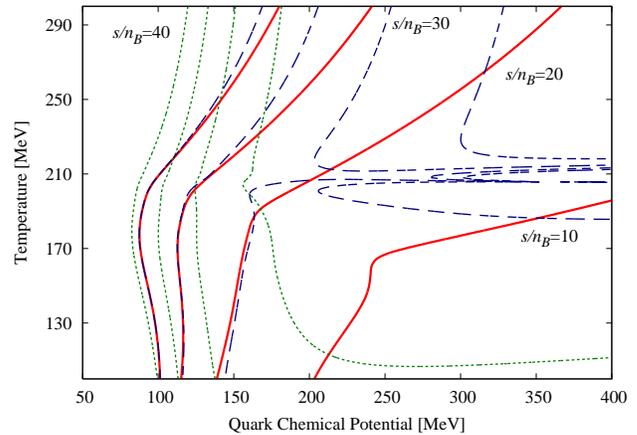}
\caption{Comparison between the full results (solid curve) and the
  truncated Taylor expansion up to the $\mu^6$-term of $s(\mu)$ and
  the $\mu^5$-term of $n(\mu)$.  The (green) dotted and (blue) dashed
  curves represent the estimates by the Taylor series fitted to
  reproduce the full results in the ranges from zero to $50\MeV$ and
  $100\MeV$, respectively.  Four curves are for $s/n_B=40$, $30$,
  $20$, $10$ from the left to the right.}
\label{fig:ex_isen_3}
\end{figure}


Figure~\ref{fig:ex_isen_2} shows the $s/n_B$-constant curves with and
without truncation on the $\mu$-$T$ plane.  The solid curves represent
the full results without truncation.  We have drawn the (green) dotted
curves by choosing $N=2$ for $s(\mu)$ and $M=1$ for $n_B(\mu)$ and the
fit range as $\mu_0=50\MeV$.  We see that the truncated series can
approximate the curve for $s/n_B=40$ well, and the curve for
$s/n_B=30$ is still close to the full estimate, but the dotted curve
for $s/n_B=20$ has a huge deviation from the corresponding solid
curve.  If we determine the Taylor expansion coefficients in a wider
range as $\mu_0=100\MeV$, as shown by the (blue) dashed curves, the
agreement becomes better, of course.  Not only the curve for
$s/n_B=40$ but also for $s/n_B=30$ agrees quite well with the full
results.  Besides, the curve for $s/n_B=20$ is significantly improved
at $T\gtrsim200\MeV$, while it does not fit the full results at lower
$T$.  It seems that the curve for $s/n_B=10$ is too far from $\mu_0$
to perceive the effect of changing $\mu_0$.

It is intriguing to increase the truncation degree of the Taylor
expansion to discuss how much the approximation is improved.  We leave
the terms up to the $\mu^6$ order in the expansion of $s(\mu)$
(i.e.\ $M=3$) and the $\mu^5$ order in the expansion of $n_B(\mu)$
(i.e.\ $N=2$) and have found the curves in Fig.~\ref{fig:ex_isen_3}.
Although the results at low $T$ become slightly better than
Fig.~\ref{fig:ex_isen_2}, it is unexpected that the agreement at high
$T$ goes worse!  This undesirable poor convergence turns out to stem
solely from the Taylor expansion of $n_B(\mu)$. 


\begin{figure}
\includegraphics[width=8.5cm]{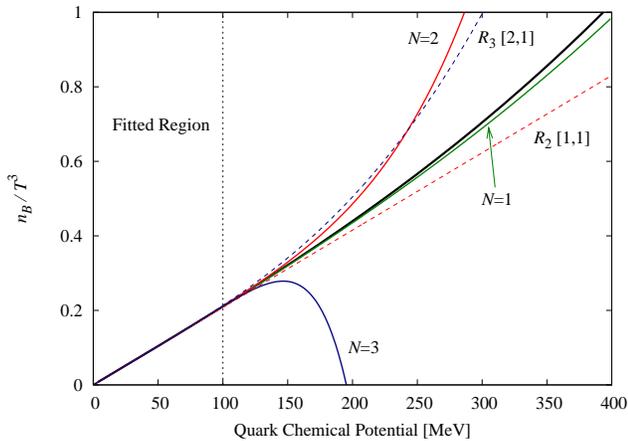}
\caption{The bold solid curve represents the full data of $n_B(\mu)$
  in the model calculation which is Taylor expanded in the $\mu$-range
  of $[0,100\MeV]$.  The thin solid curves represent the truncated
  results up to $N=1$, $N=2$, and $N=3$ as indicated by the labels.
  The dotted curves are the Pad\'{e}-improved results for $N=2$ and
  $N=3$.}
\label{fig:ex_n}
\end{figure}


To examine the problem concretely, we present a plot for $n_B(\mu)$ in
Fig.~\ref{fig:ex_n}.  The truncated results at $N=1$ (green dotted
curve) can well describe the bold solid curve which is the full data.
As we go to the higher order, however, the truncation leads to a
larger deviation from the full results, implying that the expansion
seems to fail.  If the numerical accuracy is arbitrarily good as
needed, or equivalently, $\mu_0\to\infty$, the Taylor expansion would
work with very small coefficients for the higher order terms in the
expansion series.  In practice, however, the available accuracy is
limited, and the artifact may enter, which eventually goes wrong for
$\mu$ away from the fitted region.  Here, at the same time, we should
emphasize that there is no such weird behavior observed in the
expansion of $s(\mu)$;  the higher order we take account of, the
better convergence of $s(\mu)$ we can reach, as is naturally expected.
In view of this, hence, it could be conceivable that there may be some
profound reason why only the expansion of $n_B(\mu)$ is dangerous.

We can thus learn the following important lessons from these analyses
using the model.
\vspace*{1mm}

\noindent
1) The Taylor expansion method does not work in the low-$T$ and
high-$\mu$ region.  The adiabatic path for $s/n_B\lesssim20$ deduced
from the expansion may be totally different from the true results.
This value of $s/n_B$ happens to be close to the threshold below which
the pressure starts moving apart from that at $\mu=0$ as seen in
Fig.~\ref{fig:p_isen_ns}.  This fact implies that the Taylor expansion
may well work only in the regime where the density effect is not such
appreciable.
\vspace*{1mm}

\noindent
2) Even at high $T$ where the Taylor expansion is believed to work
nicely, the expansion of the baryon density $n_B$ may be problematic
suffering from the uncertainty in the higher order terms than
$\mathcal{O}(\mu^3)$.  One of the simplest remedies for this
pathological expansion is the Pad\'{e} improvement as addressed in
Ref.~\cite{Gavai:2008zr}.  We can see in Fig.~\ref{fig:ex_n} that the
Pad\'{e}-improved results, $R_2[1,1]$ for $N=2$ and $R_3[2,1]$ for
$N=3$, certainly come close to the full curve.  [$R_2[1,1]$ denotes
  $c_1\mu(1+c_2\mu^2)/(1+c_3\mu^2)$ with $c_1$, $c_2$, $c_3$ fixed to
  yield the original series up to $N=2$ and $R_3[2,1]$ should be
  understood likewise.]   We note that the entropy density $s$ does
not have such a kind of expansion problem at all.


\section{STRANGE QUARK CHEMICAL POTENTIAL}
\label{sec:mus}

As promised in Sec.~\ref{sec:results}, the final topic discussed in
this paper is the induced chemical potential for strange quarks to
keep the strangeness neutrality.  In general the condition of $n_s=0$
requires a positive $\mu_s$.  This can be intuitively understood as
follows.

The strong interaction does not change the number of strange quarks
but can make strange particles in a process, for example, such as
$\pi^-+p\to K^0+\Lambda$ and $\pi^-+p\to K^++\Sigma^-$ etc.  The
strangeness of $\Lambda$ and $\Sigma^0$, $\Sigma^\pm$ is negative one,
meaning that they contain positive one strange quark.  Therefore, if
$p$ and $n$ are abundant at finite baryon chemical potential, the
strong interaction pushes strange quarks into strange baryons , that
results in $\mu_s>0$.  [For more phenomenological details see
Ref.~\cite{Letessier:2002gp} and references therein, and see also
Ref.~\cite{Bravina:2008ra}.]

This sort of dynamics is completely missing in the NJL model without
color confinement.  If one solves $n_s=0$ in the three-flavor NJL
model, it ends up with $\mu_s=0$, which is unphysical.  In fact a
positive $\mu_s$ has been concluded in Ref.~\cite{Bernard:2007nm} as
it should be.  In this section we will see that the PNJL model has a
crucial advantage in describing the induced $\mu_s$ correctly because
it encompasses the confinement physics.


\begin{figure}
\includegraphics[width=8.5cm]{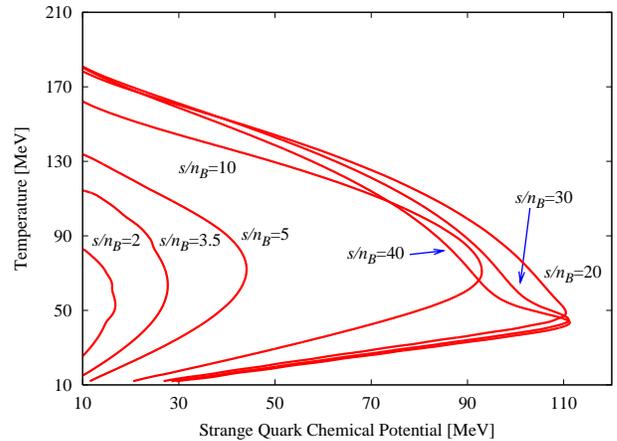}
\caption{Induced $\mu_s$ necessary to keep $n_s=0$ for various values
  of $s/n_B$ in the PNJL model calculation.}
\label{fig:isen_traj_mus}
\end{figure}


Let us first show our numerical results in
Fig.~\ref{fig:isen_traj_mus}.  A non-zero and positive $\mu_s$
certainly appears in the PNJL model unlike the NJL model.  The
numerical values are qualitatively consistent with the lattice results
in Ref.~\cite{Bernard:2007nm}, though the quantitative comparison is
not straightforward.  We would say that this demonstration of
$\mu_s>0$ adds another example to the successful PNJL model
applications besides the bulk thermodynamics.  The rest of this
section is devoted to explaining how the Polyakov loop makes it
possible to accommodate the induced $\mu_s$.

We recall that the Polyakov loop coupling takes a form of
\begin{equation}
 \ln\det\bigl[1+L\,\rme^{-(\varepsilon-\mu)/T}\bigr]
  +\ln\det\bigl[1+L^\dagger\,\rme^{-(\varepsilon+\mu)/T}\bigr] \,,
\label{eq:coupling}
\end{equation}
in each flavor sector.  As long as $\mu$ is small
Eq.~(\ref{eq:coupling}) is well approximated as
\begin{equation}
 \simeq 3\,\rme^{-\varepsilon/T}\bigl(\ell\,\rme^{\mu/T}
  +\bar{\ell}\,\rme^{-\mu/T}\bigr) \,.
\end{equation}
For light flavors at $\mu>0$, therefore, the source weight for
anti-quarks is larger than that for quarks which yields
$\bar{\ell}>\ell$, as is consistent with the argument given below
Eq.~(\ref{eq:determinant}).  It might be a bit confusing but
$\ell\,\rme^{\mu/T}$ is the source for anti-quarks because the
derivative of $\bar{\ell}\,\ell$ with respect to $\ell$ gives the
anti-Polyakov loop $\bar{\ell}$.  Then, in the heavy flavor sector,
the neutrality condition means in the same approximation,
\begin{equation}
 \frac{\partial}{\partial\mu_s} \bigl(\ell\,\rme^{\mu_s/T}
  +\bar{\ell}\,\rme^{-\mu_s/T}\bigr) = 0 \,,
\end{equation}
which immediately leads to
\begin{equation}
 \mu_s = \frac{T}{2}\log\bigl(\ellb/\ell\bigr) \,.
\label{eq:mus}
\end{equation}
Interestingly enough, the above equation~(\ref{eq:mus}) holds
approximately in the entire $\mu$-$T$ plane with only a $3\%$
violation at worst!  This is an intriguing relation discovered in a
heuristic manner in the PNJL model and could be tested in the future
lattice simulation.

One might wonder why the PNJL model could yield a positive $\mu_s$
though it does not properly describe the confined baryons such as $p$
and $n$.  As a matter of fact, at sufficiently high temperature, the
thermal system may well consist of mesons and quarks rather than
baryons, in which $\pi^-+p\to K^0+\Lambda$, for instance, should be
replaced by $\pi^-+u\to K^0+s$ in terms of quarks.  The Polyakov loop
mediates the mesonic correlation through the color average, so that
this kind of process is to be taken into account in the PNJL model.
Moreover this process via quarks is rather realistic because in the
hadronic phase the cross section of kaons is small and $n_s=0$ would
no longer hold at the later stage of evolution~\cite{Bravina:2008ra}.

From Eq.~(\ref{eq:mus}) it is easy to confirm that $\mu_s\to0$ when
$T$ grows large.  If $T$ exceeds $T_c$, both the Polyakov loop and the
anti-Polyakov loop come close to unity, and thus their ratio is nearly
one, the logarithm of which is zero.  This naturally coincides with
the intuition that deconfined quarks have no correlation and $\mu_s=0$
corresponds to $n_s=0$ just like in the NJL model.  In contrast to the
high-$T$ situation, the confined phase at small $T$ has
$\ell\simeq\bar{\ell}\simeq 0$.  The ratio of $\ell$ and $\bar{\ell}$
could be any number.  If $\bar{\ell}$ goes to zero much slower than
$\ell$, the ratio can become arbitrarily large.

In the best of our knowledge Eq.~(\ref{eq:mus}) is the very first
demonstration that the discrepancy between $\ell$ and $\bar{\ell}$ at
finite $\mu$ has a physically significant consequence.  The importance
of $\ell\neq\bar{\ell}$ has been overlooked maybe because the
difference, $\bar{\ell}-\ell$, is negligibly small (only a few $\%$ at
most) as compared to $\tfrac{1}{2}(\ell+\bar{\ell})$.  The essential
point is that we consider not the difference but the ratio,
$\bar{\ell}/\ell$, which may take a huge value when $\ell\simeq0$ and
$\bar{\ell}\simeq 0$.


\begin{figure}
\includegraphics[width=9cm]{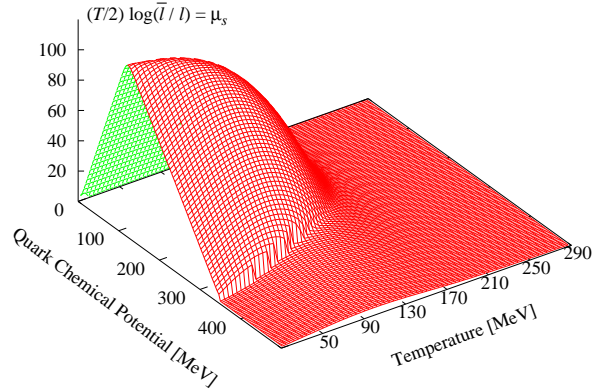}
\caption{Induced $\mu_s$ (or logarithm of the ratio between $\ell$ and
  $\ellb$) on the $\mu$-$T$ plane.}
\label{fig:diff_ratio}
\end{figure}


We plot the induced $\mu_s$ on the $\mu$-$T$ plane in
Fig.~\ref{fig:diff_ratio}.  The functional shape is quite
characteristic at small temperature.  Let us disclose the origin of
this triangle-peak structure at low $T$ separating the $\mu$-region
into three pieces:
\vspace*{2mm}

\noindent
i) Quark-regime --- At small $\mu$ up to the peak position, the
induced $\mu_s$ rises linearly along with $\mu$.  In this region the
mesonic correlation is the governing dynamics as already explained
above.
\vspace*{2mm}

\noindent
ii) Diquark-regime --- The turning point of $\mu$ where $\mu_s$ starts
decreasing corresponds to the chemical potential with which the
diquark excitation $\bar{\ell}\,\rme^{-2(\varepsilon-\mu)/T}$ is
energetically more favorable than the quark excitation
$\ell\,\rme^{-(\varepsilon-\mu)/T}$.  Using Eq.~(\ref{eq:mus}) we can
derive the condition $\mu>\tfrac{1}{3}\varepsilon$ for the diquark
excitation overcoming the quark excitation.  Since the constituent
quark mass is $336\MeV$ in this model, the threshold should be
$\mu\sim 110\MeV$.  This estimate is really consistent with our
numerical results shown in Fig.~\ref{fig:diff_ratio}.  In this
$\mu$-region $\mu_s$ decreases because diquarks behave like
anti-quarks in color space.  In other words the mesonic
(quark--anti-quark) correlation is taken over by the baryonic
(quark--diquark) correlation which carries nonzero baryon number and
so $\mu_s$ is partially canceled by this effect.
\vspace*{2mm}

\noindent
iii) Baryonic-regime --- If $\mu$ is greater than $\varepsilon$ the
color singlet contribution, $\rme^{-3(\epsilon-\mu)/T}$, is dominant,
which is interpreted as the baryonic excitation.  In this regime the
Polyakov loop is decoupled from the dynamics and there is no
confinement effect, even though the Polyakov loop is zero.  (Such a
state is recently named the
``quarkyonic phase''~\cite{McLerran:2007qj,Glozman:2007tv,%
Miura:2008gd,McLerran:2008ua}.)  Then $\mu_s=0$ suffices for $n_s=0$.


\section{SUMMARY}
\label{sec:summary}

We have calculated the isentropic trajectories on the phase diagram in
the $\mu$-$T$ plane using the PNJL model.  Our results are
quantitatively consistent with the lattice results in the high-$T$ and
low-$\mu$ region where the lattice data is available by means of the
Taylor expansion.

To test whether the Taylor expansion is under theoretical control or
not, we have expanded our numerical results of the entropy $s$ and the
baryon density $n_B$ into polynomial series in terms of $\mu$.  We
have confirmed that the Taylor expansion can access the trajectories for
$s/n_B=40$, $30$, and $20$, but cannot for $s/n_B=10$ at which the
isentropic thermodynamics differs substantially from that at zero
density.  We have also realized that $n_B(\mu)$ has pathological
behavior if expanded at high temperature, which can be cure by the
Pad\'{e} approximation.

Finally we have discussed the induced strange quark chemical potential
$\mu_s$ to keep $n_s=0$ in the system ruled by the strong
interaction.  We have found the interesting relation between $\mu_s$
and the ratio of the Polyakov loop $\ell$ and the anti-Polyakov loop
$\bar{\ell}$.  This prediction could be confirmed in the lattice QCD
simulation.

In this work we neglected the effect of the soft-mode fluctuations
around the critical point.  This is one important direction of the
future extension.  Another interesting direction is the origin of the
poor convergence in the expansion of $n_B(\mu)$.  This is not fully
understood in the present work;  other thermodynamic quantities like
$s(\mu)$ have a smooth expansion but only $n_B(\mu)$ fails.  It would
be also interesting, if the baryon number susceptibility does not have
such a problem of expansion (we guess so), to validate the idea that
the QCD critical point is to be deduced from the radius of convergence
using the model.  Although these issues are all beyond our current
scope, we believe that the present research contributes to opening
these extensions.


\acknowledgments
The author thanks H.~Iida, L.~Levkova, T.~T.~Takahashi for
discussions.  He thanks the Institute for Nuclear Theory at the
University of Washington for its hospitality and the Department of
Energy for partial support during the completion of this work.  He is
supported by Japanese MEXT grant No.\ 20740134 and also supported in
part by Yukawa International Program for Quark Hadron Sciences.


\end{document}